\documentclass{aa}  

\usepackage{graphicx, subfigure}
\usepackage{natbib}
\usepackage{verbatim}
\usepackage{txfonts,epsfig,url,twoopt}
\usepackage[breaklinks=true]{hyperref}
\hypersetup{colorlinks=true,citecolor=blue}
\bibpunct{(}{)}{;}{a}{}{,} % to follow the A&A style
\usepackage{hyperref}
\usepackage{color} 
\begin{document}

   \title{How much does turbulence change the pebble isolation mass for planet formation?}
   \author{S.~Ataiee \inst{1}
          \and
          C.~Baruteau \inst{2}
          \and
          Y.~Alibert \inst{1}
          \and
          W.~Benz \inst{1}
          }

   \institute{University of Bern, Physics Institute, Space Research and Planetary Sciences, Sidlerstrasse 5, CH-3012 Bern, Switzerland\\
              \email{sareh.ataiee@space.unibe.ch}
         \and
            IRAP, Universit{\'e} de Toulouse, CNRS, UPS, Toulouse, France
             }

   \date{}

% \abstract{}{}{}{}{} 
% 5 {} token are mandatory
 
  \abstract
  {When a planet becomes massive enough, it gradually carves a partial gap around its orbit in the protoplanetary disk. A pressure maximum can be formed outside the gap where solids that are loosely coupled to the gas, typically in the pebble size range, can be trapped. The minimum planet mass for building such a trap, which is called the pebble isolation mass (PIM), is important for two reasons: it marks the end of planetary growth by pebble accretion, and the trapped dust forms a ring that may be observed with millimetre observations.}
  {We study the effect of disk turbulence on the pebble isolation mass and find its dependence on the gas turbulent viscosity, aspect ratio, and particles Stokes number.}
  {By means of 2D gas hydrodynamical simulations, we found the minimum planet mass to form a radial pressure maximum beyond the orbit of the planet, which is the necessary condition to trap pebbles. We then carried out 2D gas plus dust hydrodynamical simulations to examine how dust turbulent diffusion impacts particles trapping at the pressure maximum. We finally provide a semi-analytical calculation of the PIM based on comparing the radial drift velocity of solids and the root mean square turbulent velocity fluctuations around the pressure maximum.}
  {
    From our results of gas simulations, we provide an expression for the PIM versus disk aspect ratio and turbulent viscosity. Our gas plus dust simulations show that the effective PIM can be nearly an order of magnitude larger in high-viscosity disks because turbulence diffuse particles out of the pressure maximum. This is quantified by our semi-analytical calculation, which gives an explicit dependence of the PIM with Stokes number of particles.
  }
   {
     Disk turbulence can significantly alter the PIM, depending on the level of turbulence in regions of planet formation.
   }


   \maketitle
%
%-------------------------------------------------------------------

   % ===============
   \section{Introduction} 
   % ===============
   \label{intro}
   Pebbles -- solids in the mm-cm size range -- are important ingredients for planet formation in protoplanetary disks as they can efficiently speed up the formation of planetary cores \citep[e.g.][]{2010A&A...520A..43O,2010MNRAS.404..475J,2012A&A...544A..32L}. Solids couple more or less to the disk gas depending on their size, internal density, and on gas density and temperature. The accretion of well-coupled solids on a planetary core is inefficient as they are fully dragged by the gas across the orbit of the planet. The accretion of decoupled solids is also inefficient because of their high velocity relative to the planet. But marginally coupled solids can be efficiently accreted as the drag force around the planetary core notably reduces their velocity relative to the planet \citep{2010A&A...520A..43O}. This is the case for solids in the pebble size range in typical regions of planet formation.
 
   Another interesting aspect of pebbles is their potential observational signatures in the (sub)mm continuum emission of protoplanetary disks, in particular when they are trapped at pressure maxima. These pebbles can form bright and dark emission rings when there is a radial pressure bump \citep[e.g.][]{Andrews16TWHya} or lopsided emission structures if there is an azimuthal pressure bump such as a vortex in the disk gas \citep[e.g.][]{Fuente17}. Observing the cold dust emission of disks can give constraints about the physical process behind the dust trapping in these pressure bumps such as the possible existence of a planet \citep{2012MNRAS.419.1701R,2016ApJ...818..158A}.
   
   The pebble isolation mass (hereafter PIM) is defined as the minimum planetary mass that prevents pebbles from drifting to and being accreted by a planet in a protoplanetary disk. As the planet grows, the spiral wakes that it generates in the protoplanetary disk progressively become shocks closer to the planet. This process leads to the opening of a partial gap in the gas disk about the orbital radius of the planet \citep{Rafikov02}. A pressure maximum builds up at the outer edge of the gap of the planet, where pebbles should be trapped \citep[e.g.][]{2003ApJ...583..996H,2012A&A...545A..81P}. This trapping may have two major implications: the growth of the planet by pebble accretion ceases and the trapped pebbles can produce ring-like structures in the (sub-)mm continuum emission. Said differently, the PIM can be seen as the maximum mass of a planetary core for \textit{in-situ} pebble-mediated planet formation, and as the minimum planetary mass that may produce observable ring-like features in observations of the radio continuum.
  
   An accurate determination of the PIM, particularly its functional dependence on disk parameters, is especially important for global models of planet formation and evolution. \cite{2014A&A...572A..35L} investigated the dependency of the PIM on the scale height of the disk and location of the planet, but not on turbulent viscosity of the disk. However, the formation of even a partial gap in the disk gas around the orbit of the planet is sensitive to the turbulent viscosity of the disk, as shown for instance in the gap-opening criterion formulated in \citet{2006Icar..181..587C}. Furthermore, the presence of a pressure maximum is a necessary condition for trapping solids that are marginally coupled to the gas beyond the orbit of the planet, but it is not sufficient. Turbulent diffusion also acts on solids and may kick them out of the equilibrium location defined by the pressure maximum. This depends on how strong turbulent diffusion is compared to the radial drift due to gas drag. Because turbulent diffusion depends both on the turbulent viscosity of the disk and the size of the solids via their stopping time \citep{2007Icar..192..588Y}, the PIM should therefore also depend on solids size. We note that very small particles, which are well coupled to the gas, are not trapped at the pressure maximum \citep[e.g.][]{2012ApJ...755....6Z} unless perhaps the planet is so massive that it keeps the gas outside the  orbit of the planet from flowing across the gap.
   
   The aim of this study is to examine how the turbulent viscosity of the disk impacts the PIM with a combination of hydrodynamical simulations and semi-analytical calculations. The manuscript is organized as follows: In Sect.~\ref{simus} we describe our physical model and numerical set-up, and we present the results of simulations. Our gas plus dust simulations show that the inclusion of dust turbulent diffusion can increase the PIM by nearly an order of magnitude at high gas turbulent viscosities. In Sect.~\ref{semiformula} we derive a semi-analytical expression for the PIM that is valid for a broad range of disk scale heights, viscosities, and pebble sizes, and we discuss its accuracy. This expression can easily be used in population synthesis models of planet formation and evolution, or for the interpretation of radio disk observations. We finally summarize our results in Sect.~\ref{summary}.

   % ===============
   \section{Hydrodynamical simulations} 
   % ===============
   \label{simus}
   The PIM is {\it a priori} a function of the disk's scale height $H$, its turbulent kinematic viscosity $\nu$, and the particles stopping time $t_{\rm s}$. These quantities can be expressed as dimensionless parameters that are the disk's aspect ratio $h$, $\alpha$ turbulent viscosity, and Stokes number $\mathit{St}$ via
   \begin{equation}
   h = \frac{H}{r},\hspace{1cm} \alpha = \frac{\nu}{c_{\rm s} H},\hspace{1cm} \mathit{St}=\frac{t_{\rm s}}{t_{\rm eddy}},
   \label{dimendionless}
   \end{equation}
   where $r$ denotes the distance from the central star and $c_{\rm s}$ is the sound speed. The Stokes number in the disk midplane is defined as the ratio between the particles stopping time $t_{\rm s}$ (the time particles need to adjust their motion to the gas because of drag forces), and the eddy turnover time $t_{\rm eddy}$, that is the correlation timescale of turbulent fluctuations, which is typically of the order of the orbital timescale (see e.g. \citealp{2007Icar..192..588Y}). For particle sizes much smaller than the molecular mean-free path (Epstein regime), which is the case in our semi-analytic calculation and our 2D simulations, the stopping time is $t_{s}=\pi \rho_{\rm s} s / 2 \Sigma \Omega $ with $\rho_{\rm s}$ the particles internal density, $s$ their physical radius, $\Sigma$ and $\Omega$ the surface density and the angular velocity of the gas at the location of the particles  \citep{1977MNRAS.180...57W}. Adopting $t_{\rm eddy} = \Omega^{-1}$ results in $\mathit{St}=\pi s \rho_{\rm s}/2 \Sigma$.
   
   To examine how disk turbulence impacts the PIM, we adopted a two-step strategy. In the first step, we carried out series of 2D gas hydrodynamical simulations of disk-planet interactions for a range of values for $h$ and $\alpha$. More specifically, for each pair of $\{h,\alpha\}$, we found the minimum planet mass for which the gas azimuthal velocity ($v_{\varphi}$) can reach or slightly exceed the Keplerian velocity ($v_{\rm K}$) beyond the orbital radius of the planet due to the local deposition of angular momentum by the outer wake of the planet. This is the necessary condition for pebbles to be trapped beyond the orbit of the planet. We provide a fitting formula for this PIM as a function of $h$ and $\alpha$. In this step, the turbulent viscosity modifies the gas velocities through the gas momentum equation and results in different PIM for different viscosities. That means that in this step we investigated the indirect effect of  turbulent viscosity on  particle trapping. In the second step, we carried out a few 2D gas plus dust hydrodynamical simulations of disk-planet interactions with a range of dust sizes to investigate how much the turbulent diffusion of dust alters the PIM as inferred from gas-only simulations. In this step, we examined the direct influence of viscous turbulent diffusion on the pebble trapping in addition to the indirect effect that is measured in the first step. These gas plus dust simulations are carried out for a specific value of $h$, but for a range of $\alpha$ values. After describing the physical model and numerical set-up of our simulations in Sect.~\ref{modelandsetup}, we present the results of the gas simulations in Sect.~\ref{gassims} and those of the gas plus dust simulations in Sect.~\ref{dustsimsresults}.
 
  % -------------------
   \subsection{Physical model and numerical set-up} 
   \label{modelandsetup}
   % -------------------
   We carried out our gas and gas plus dust hydrodynamical simulations with the 
   code Dusty FARGO-ADSG. It is an extended version of the public code 
   \href{http://fargo.in2p3.fr/-FARGO-ADSG-}{FARGO-ADSG}
\citep{2000A&AS..141..165M, Baruteau2008a, Baruteau2008b}, which includes dust \citep{2016MNRAS.458.3927B}. 
\\
\par \noindent {\it Gas --} The gas continuity and momentum equations 
are solved in a 2D grid with polar coordinates $\{r, \varphi\}$. 
We use 666 grid cells uniformly spaced 
in radius between 0.5$a$ and 2.5$a$ ($a$ denotes the semi-major axis of the planet), and 600 cells evenly spaced in azimuth between 0 and 2$\pi$. 
The initial surface density profile of the gas is $\Sigma_{0}(r)=10^{-4} (M_{\star}/a^2) \times (r/a)^{-1/2}$ with $M_{\star}$ the mass of the central star. This corresponds to a gas surface density $\sim 900$ g cm$^{-2}$ at 1~au for a solar-mass star. The Toomre-Q parameter never falls below $\approx$25 throughout our series of simulations, and for this reason gas self-gravity is discarded. Furthermore, for simplicity we adopt a locally isothermal equation of 
state with the gas 
pressure given by $P = \Sigma c_{s}^2$, where the sound speed (or 
temperature) are fixed in time. 
The disk aspect ratio, $h=c_{\rm s}/v_{\rm K}$, is assumed to be uniform, which translates 
to a temperature profile $T \propto r^{-1}$. The aspect ratio $h$ is varied 
from 0.03 to 0.06 by steps of $5\times10^{-3}$. The $\alpha$ turbulent viscosities 
taken in this work are $5\times10^{-4}, 10^{-3}, 2\times10^{-3}, 5\times10^{-3}$, and $10^{-2}$. This range of $h$ and $\alpha$ values are meant to reflect the range of 
temperatures and rates of turbulent transport expected in the inner regions 
($r \lesssim 20 AU$) of protoplanetary disks \citep[e.g.][]{2015MNRAS.454.1117S}.
\\
\par \noindent {\it Dust --} When dust is included in the simulations, it is modelled as Lagrangian test
 particles that feel the gravity of the star and planet, as well as the gas drag. 
 We do not consider the dust back-reaction on the gas in this study. This effect might be important in long-term studies as we briefly discuss in Sec.~\ref{summary}. Turbulence is modelled by applying stochastic kicks to the particles position following the method in \citet{2011ApJ...737...33C}, in which the spatial dependence of the turbulent diffusion of the dust is discarded, however. Specifically, the radial kicks have mean $\langle\delta r\rangle = D_{\rm d}\,dt\,\partial_{r}\Sigma/\Sigma$ and standard 
 deviation $\sigma_r = \sqrt{2 D_{\rm d} dt}$, and the azimuthal kicks have mean  $\langle\delta\varphi\rangle = D_{\rm d}\,dt\,\partial_{\varphi}\Sigma/(r^2\Sigma)$ and standard 
 deviation $\sigma_{\varphi} = \sigma_r /r$. For the dust's turbulent diffusion $D_{\rm d}$, we use $D_{\rm d}=\nu (1+4\mathit{St}^2)/(1+\mathit{St}^2)^2$ from \citep{2007Icar..192..588Y}. When the particles stopping time becomes shorter than 
 the hydrodynamical timestep set by the gas CFL condition in the simulations, 
 the so-called short friction time approximation is used to update particles velocities \citep{2005ApJ...634.1353J}. In this work we use 50000 dust 
 particles with a size distribution $n(s) \propto s^{-1}$ for sizes $s$ between 
 1 mm and 300 mm. The particles internal density is set to 2 g cm$^{-3}$. These particle sizes and internal density correspond to initial Stokes numbers $\mathit{St} \in [3 \times 10^{-4}-0.2]$. The Stokes number of the particles changes as they drift or diffuse in the disk inversely proportionally to the gas surface density at the particles location. The size range of the particles is chosen such that, first, there are enough particles with $\mathit{St} > 0.05$ that are trapped at the edge of the gap of the planet and, second,~there are still particles with $\mathit{St}<0.05$ that are not trapped at the gap edge and drift to the inner disk. The $\mathit{St} \approx 0.05$ threshold is meant to be a slightly lower bound for the typical range of Stokes numbers for which pebble accretion is most efficient, which is $\mathit{St}\in [0.1-1]$ \citep[see e.g.][]{BBMM2016}.  Said differently, since this study deals with the PIM, we do not investigate the trapping of particles with  a smaller Stokes number, as they should not contribute significantly to pebble accretion \citep[e.g][]{2014A&A...572A.107L}. However, one can estimate the PIM for smaller Stokes numbers from the semi-analytical formula presented in Sect.~\ref{semiformula}.
\\
\par \noindent {\it Planet --}
The planet is held on a fixed circular orbit in all our simulations, and its gravitational 
 potential is smoothed over a softening parameter $\epsilon=0.4H(a)$. Its mass is gradually increased over the first five orbits in the simulations to let the gas adjust to the insertion of the planet in the computational grid. 
 To avoid reflections of the planet wakes near to the radial edges of the grid,
so-called wave-killing zones are used in which gas fields are damped
towards their initial radial profile \citep{2006MNRAS.370..529D}. All 
calculations are carried out in a frame co-rotating with the planet. 
\\
\par \noindent {\it Units and running times --}
Results of simulations are expressed in the following units: the mass unit 
is the mass of the central star ($M_{\star}$), the length unit is the planet's 
(fixed) semi-major axis ($a$), and the time unit is $\Omega_{\rm K}^{-1}(a)$. We denote by $q$ the 
planet-to-star mass ratio and by $q_{\oplus}$ the Earth-to-Sun mass ratio ($q_{\oplus} \approx 3 \times 10^{-6}$). 
The simulations with only gas, which are presented in Sect.~\ref{gassims}, are run over $t_{\rm run}=5000, 4000, 3000, 2000$ or $1000$~planet orbits for our models with lowest to highest viscosity. These running times are chosen such that the disk profiles reach a steady state. Simulations with gas plus dust, which are presented in Sect.~\ref{dustsimsresults}, are restarts of the gas-only simulations, where dust particles are inserted between $r=1.4a$ and $1.41a$ uniformly in radius and azimuth.

   % -------------------
   \subsection{Results of gas-only simulations: Minimum planet mass to form a local pressure bump beyond the planet's orbit} 
   \label{gassims}
   % -------------------
   % FFFFFFFFFFFF
   \begin{figure}
     \centering
     \includegraphics[width=\hsize]{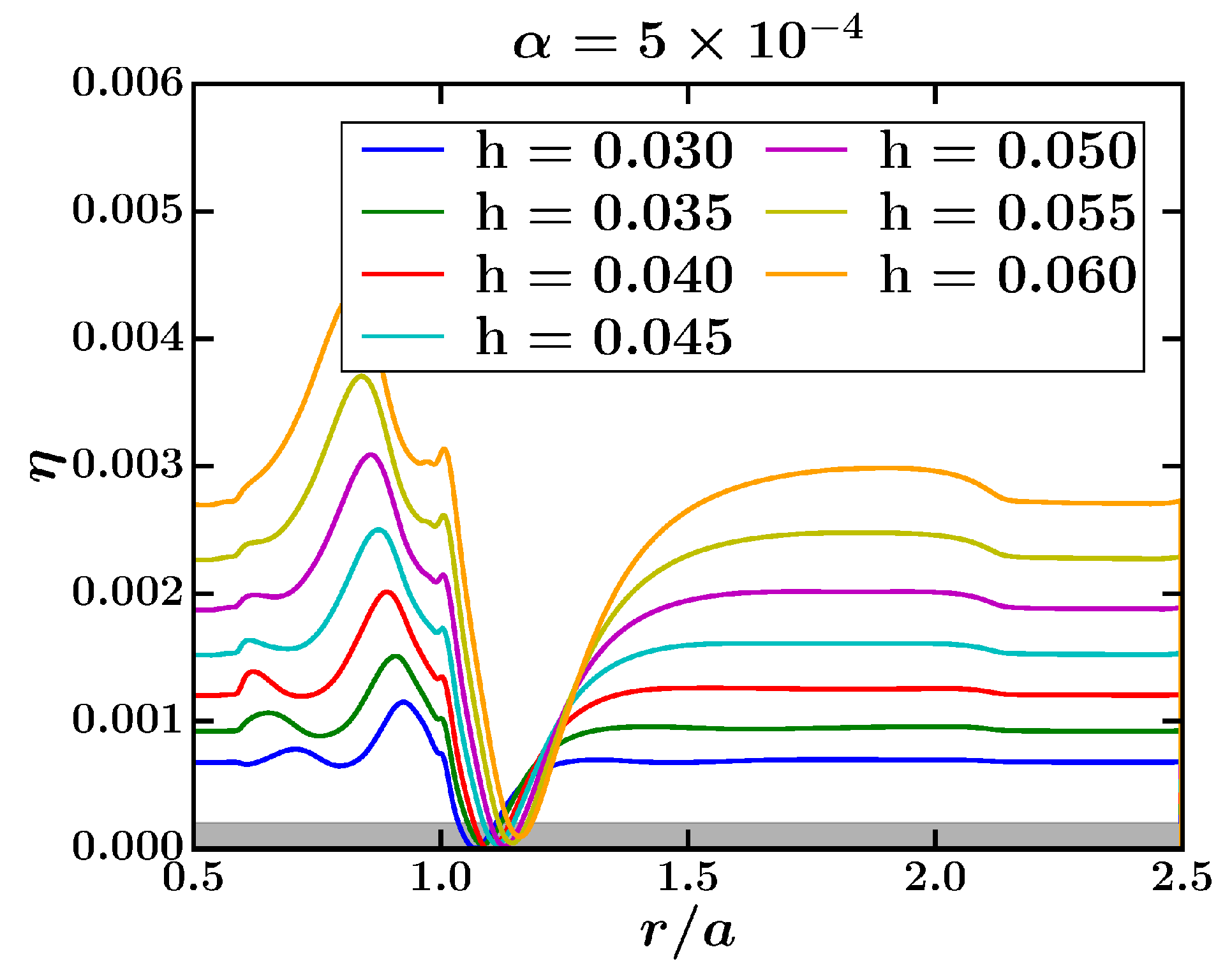}
     \caption{Illustration of the determination of the PIM via gas-only hydrodynamical simulations, where we find the minimum planet-to-star mass ratio ($q_{\rm g}$) for the formation of a pressure maximum beyond the orbit of the planet. The quantity $\eta=-h^2 \partial\log P/\partial\log r$ (averaged over azimuth) is shown as a function of orbital radius $r$ (normalized to the planet's semi-major axis $a$). Each curve is for a pair 
     $\{q_{\rm g}, h\}$ at $\alpha=5 \times 10^{-4}$. The shaded black area shows where 
     $\eta \approx 0$ for the practical determination of $q_{\rm g}$.}
           \label{gaseta}
    \end{figure}
    % FFFFFFFFFFFF   

   Our gas hydrodynamical simulations were carried out to find the minimum planet mass for which a radial pressure maximum forms beyond the orbit of the planet as a function of disk aspect ratio $h$ and $\alpha$ turbulent viscosity. As already mentioned at the beginning of Sect.~\ref{simus}, this was carried out by looking for the minimum planet-to-star mass ratio ($q_{\rm g}$) for which the gas azimuthal velocity slightly exceeds the Keplerian velocity beyond the planet's orbit, for each pair of $\{h,\alpha\}$ ;the subscript g in $q_{\rm g}$ indicates that this is the PIM inferred from the gas-only simulations without considering the effect of turbulent diffusion of the dust. Equivalently, since $v_{\varphi} = v_{\rm K} \sqrt{1-\eta}$ 
 with $\eta = -h^2 \partial\log P/\partial\log r$, another way of looking for $q_{\rm g}$ is to find 
 the planet-to-star mass ratio from which $\eta$ cancels out beyond the orbit of the planet. 
 Fig.~\ref{gaseta} shows the azimuthally-averaged radial profile of $\eta$ for 
 $\alpha = 5 \times 10^{-4}$, obtained when the planet mass reaches the PIM for 
 different values of the disk's aspect ratio $h$. As can be seen in the figure, $\eta > 0$ 
 everywhere in the disk except at the location of the radial pressure maximum, where 
 $\eta \approx 0$.

   % FFFFFFFFFFFF
   \begin{figure}
     \centering
     \includegraphics[width=\hsize]{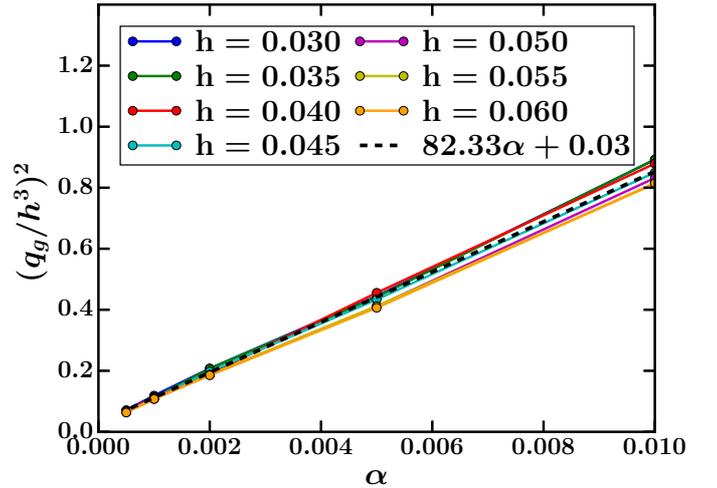}
     \caption{Calculation of the PIM via gas-only hydrodynamical simulations ($q_{\rm g}$, PIM normalized to the star mass). Results are shown for the range of disk aspect ratios $h$ and $\alpha$ turbulent viscosities considered in this work. The blue dashed curve shows the best fit for $(q_{\rm g}/h^3)^2$ vs. $\alpha$, given by Eq.~(\ref{gasfitformula}).}
           \label{gasfitsims}
    \end{figure}
    % FFFFFFFFFFFF   
    
     The values of $q_{\rm g}$ obtained for all of our models are shown in Fig.~\ref{gasfitsims}. The quantity $(q_{g}/h^3)^2$ is displayed as a function of $\alpha$ to highlight a linear relationship between both quantities. We find the best agreement for the following expression:
   \begin{equation} \label{gasfitformula}
      \left(\frac{q_{\rm g}}{h^3} \right)^2 \approx 82.33 \alpha + 0.03.
   \end{equation}
   Although this relationship shows that the PIM has a much stronger dependency on $h$ than $\alpha$, the gas turbulent viscosity can notably affect the PIM since $\alpha$ can take a large range of values in regions of planet formation. For example,  in a disk with $h=0.04$ near to the location of the planet, the PIM increases from $\sim$6 to $\sim$20 Earth masses when $\alpha$ increases from $5\times 10^{-4}$ to $10^{-2}$. 
   
\begin{figure}%[!h]
      \centering
       \includegraphics[width=\hsize]{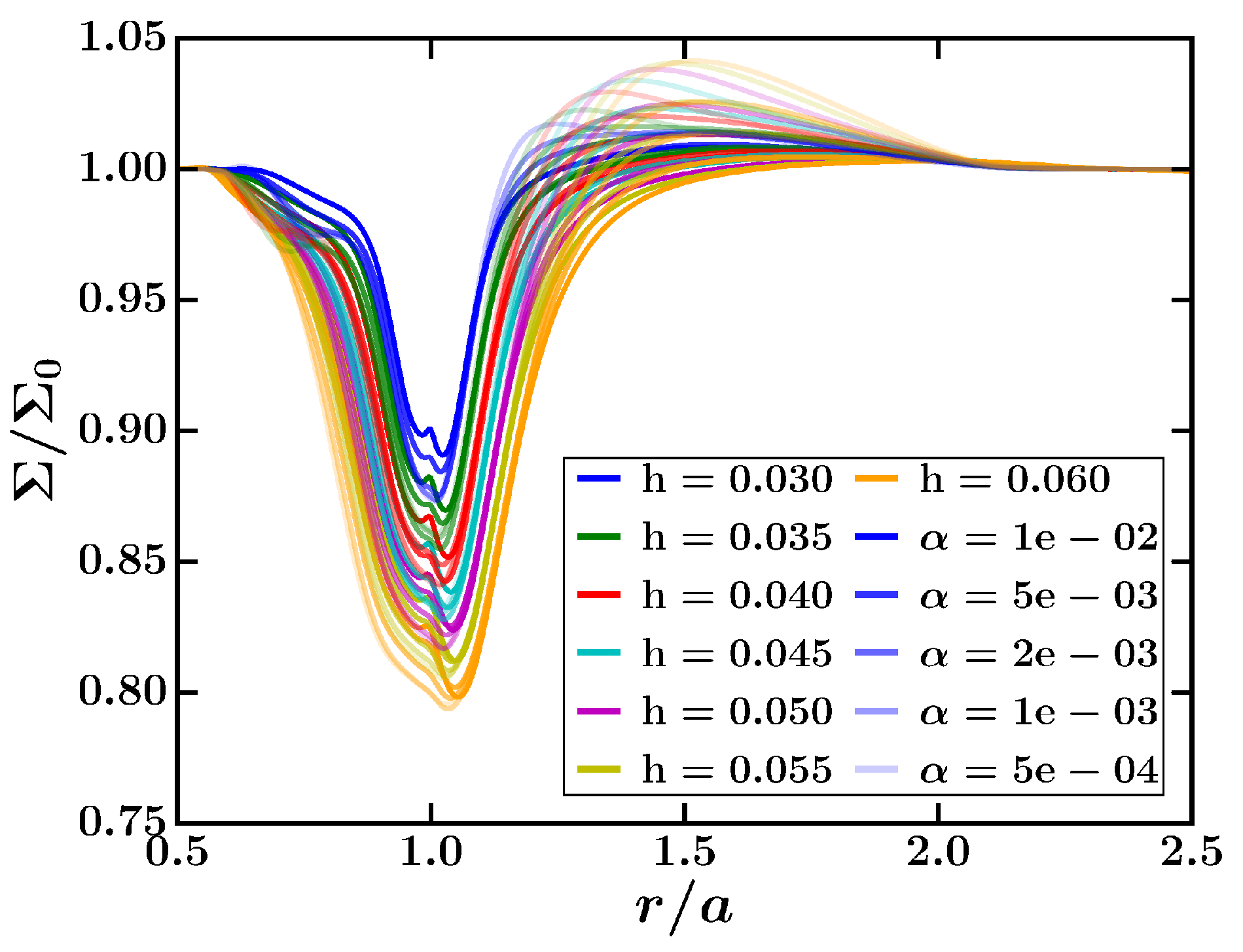}
           \caption{Gas surface density profile in all the gas-only hydrodynamical simulations where the planet takes its PIM. Line transparency varies with the $\alpha$ turbulent viscosity, and colour varies with the disk aspect ratio (see legend).}
           \label{Cridaplot}
      \end{figure}

    Eq.~\ref{gasfitformula} can be regarded as the condition for the opening of a partial gap in the gas around the orbit of the planet. This certainly differs from the gap-opening criterion formulated in \cite{2006Icar..181..587C}, which assumes planets carve a gap with a gas density drop of about 90\%. The minimum planet-to-star mass ratio ($q_{\rm gap}$) that satisfies the gap-opening criterion of \cite{2006Icar..181..587C},
    \begin{equation} \label{crida}
       \frac{3}{4} \frac{h}{(q/3)^{1/3}} + \frac{50 \alpha h^2}{q } = 1
    \end{equation}        
is given by Eq.~(10) in \citet{BaruteauPP6}, which is written as
   \begin{equation} \label{qgap}
     q_{\rm gap} = 100 \alpha h^2 \left[ \left(\sqrt{1+\frac{3h}{800 \alpha}}+1 \right)^{1/3} - \left(\sqrt{1+\frac{3h}{800 \alpha}}-1 \right)^{1/3} \right]^{-3}.
   \end{equation}
We find that $q_{\rm g}/q_{\rm gap}$ varies from about 3\% to 12\% for our range values of $h$ and $\alpha$. This point is further accentuated in Fig.~\ref{Cridaplot}, which shows that the gas surface density depletion in our models is between 10-20\%. From this we can take away that a planet that builds up a 20\% drop in the gas surface density about its orbit has reached the PIM.
   % FFFFFFFFFFFF
   % FFFFFFFFFFFF
   
   We point out that Eq.~\ref{gasfitformula} implies small PIMs for low-viscosity disks. According to Eq.~\ref{gasfitformula}, for $\alpha=0$, the PIM ranges from $\sim$1.5 Earth masses for $h=0.03$ to $\sim$12.5 Earth masses for $h=0.06$. This is consistent with the results of \citet{2013ApJ...769...41D} that basically every non-migrating planet is able to carve a partial gap or dip about its orbit in an inviscid disk.

   % -------------------
   \subsection{Results of gas plus dust simulations: Importance of turbulent diffusion on dust trapping at the pressure bump}
   \label{dustsimsresults}
   % -------------------
   
   % FFFFFFFFFFFF
   \begin{figure}
     \centering
     \includegraphics[width=\hsize]{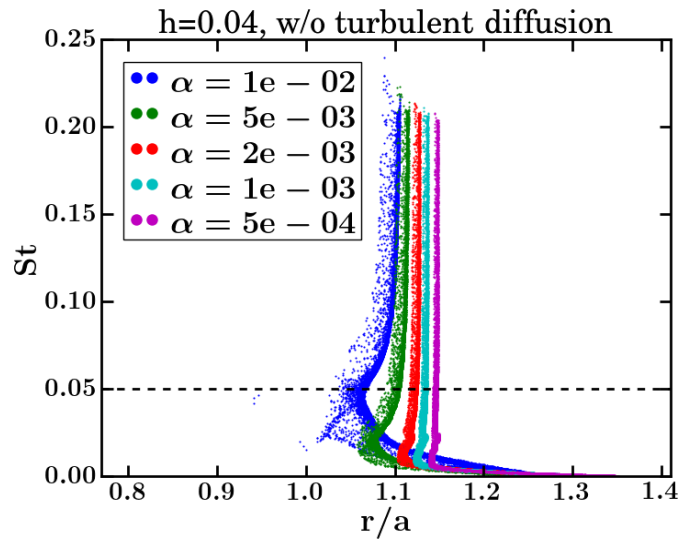}
     \caption{Results of gas plus dust hydrodynamical simulations without dust turbulent diffusion, which show that a planet with a planet-to-star mass ratio $q=q_{\rm g}$ (the normalized PIM that is inferred from our gas-only simulations) does trap solid particles with $\mathit{St}>0.05$ at the pressure maximum beyond its orbit.}
           \label{gas-dust-qg}
    \end{figure}
    % FFFFFFFFFFFF   
In the previous section, we used gas-only hydrodynamical simulations to obtain the minimum planet-to-star mass ratio ($q_{\rm g}$) for a planet to form a pressure maximum beyond its orbit, which is a necessary condition for trapping pebbles. To ensure that planets with $q=q_{\rm g}$ trap pebbles in the absence of dust turbulent diffusion, we performed several gas plus dust simulations without dust turbulent diffusion for $h=0.04$ and all our $\alpha$ values. Fig.~\ref{gas-dust-qg} demonstrates the trapping of particles with $\mathit{St} \gtrsim 0.05$ at the pressure maximum beyond the orbit of the planet, as expected. We note that the trapping location increases with decreasing the alpha turbulent viscosity, which is due to planets carving wider gaps in less viscous disks \citep[e.g.][]{2006Icar..181..587C}. The particles with $\mathit{St} \ll 0.05$ are well coupled to the gas and have not had enough time to drift significantly over the simulation time.
   
   While the formation of a radial pressure bump is a necessary condition to trap pebbles beyond the orbit of the planet, it is not sufficient. We also needed to make sure that the turbulent diffusion of the dust does not diffuse particles out of the pressure bump. Because of our large parameter space, finding the PIM via gas plus dust hydrodynamical simulations for every single set of $\{h,\alpha,\mathit{St}\}$ would be computationally expensive. We thus chose to focus on one particular aspect ratio, $h=0.04$, and find the minimum planet-to-star mass ratio that effectively traps dust particles with $\mathit{St}>0.05$ at the pressure bump. Those particles have a size $s>10\ \mathrm{mm}$ for all our $\alpha$ turbulent viscosities. As already argued in Sect.~\ref{modelandsetup}, the $\mathit{St}=0.05$ threshold can be seen as a lower bound for the typical range of particle Stokes numbers for which pebble accretion is efficient.
   
   %FFFFFFFFFFFFFFFFFFFF
    \begin{figure*}
      \centering
       \includegraphics[width=\hsize]{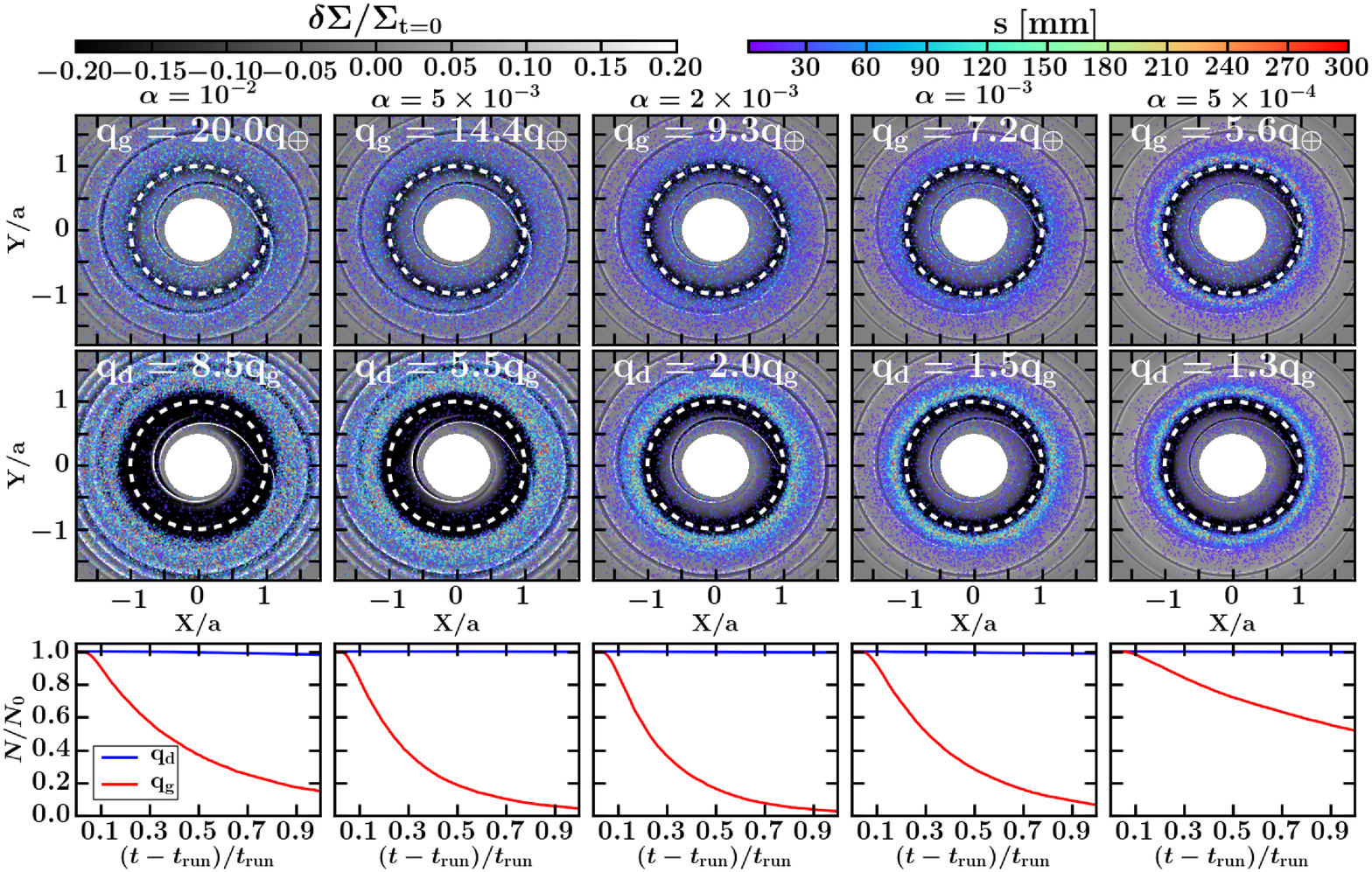}
           \caption{Results of the gas plus dust hydrodynamical simulations with dust turbulent diffusion} for $h=0.04$. The alpha turbulent viscosity decreases from left to right. First and second rows show the relative perturbation of the gas surface density (in black and white) with dots showing the location of dust particles (the colour of the dots indicates particles size). The first row shows results for $q = q_{\rm g}$, that is when the mass of the planet equals the PIM inferred from the gas-only simulations ($q_{\oplus}$ is the Earth to Sun mass ratio). The second row represents $q = q_{\rm d}$, that is when the mass of the planet equals the PIM obtained from the gas plus dust simulations. In each panel the white dashed circle indicates the orbital radius of the planet. The third row of panels represent the time evolution of $N/N_{0}$, which is the ratio of particles number with size $>10\mathrm{mm}$ that remain outside of the orbital radius of the planet.
           \label{dust2d}
      \end{figure*}   
      %FFFFFFFFFFFFFFFFFFF
   For each value of $\alpha$, we start from $q=q_{\rm g}$ (the PIM as inferred from the gas-only simulations presented in Sect.~\ref{gassims}) and increase the planet mass until dust trapping for the aforementioned sizes is observed for the whole running time of our gas+dust simulations. The corresponding planet-to-star mass ratio is denoted by $q_{\rm d}$, where the {\it d} subscript highlights that it is the mass ratio of PIM to the central star as inferred from gas plus dust simulations. Dust trapping can be quantified by counting the number of particles ($N$) with size $s>10\ \mathrm{mm}$ that remain beyond the orbital radius of the planet over time. We denote by $N_{0}$ the number of such particles at the beginning of the gas+dust simulations (at time $t=t_{\rm run}$). If the ratio $N/N_{0}$ stays close to unity for the whole running time of the gas plus dust simulations, the planet has then reached its actual PIM.
   
      The results of our gas plus dust simulations are summarized in Fig.~\ref{dust2d}. The first two rows of panels show the location of the dust particles overplotted on the relative perturbation of the gas surface density. The first row indicates $q=q_{\rm g}$ and the  second row indicates  $q=q_{\rm d}$. The alpha turbulent viscosity decreases from left to right throughout the panels. The comparison between these first two rows of panels makes it clear that for $q=q_{\rm g}$, most of the large particles have diffused out of the pressure bump and have mainly drifted inwards towards the star. This is further highlighted in the third row of panels, which represents the time evolution of $N/N_{0}$. Clearly for the smallest viscosities, $q_{\rm d}$ and $q_{\rm g}$ are in close agreement, but the discrepancy significantly increases with increasing viscosity. For instance, for $\alpha=10^{-2}$, the PIM obtained from the gas plus dust simulations is about 8.5 times larger than when obtained from the gas-only simulations. It is actually about half a Jupiter mass in the former case, and about 20 Earth masses in the latter case for a solar mass central star. 
   
   We note that the above values of $q_{\rm d}$ are for the trapping of $\mathit{St}>0.05$ particles at the pressure bump. A different (in particular smaller) threshold value should result in a different value for $q_{\rm d}$. Instead of finding the Stokes- or size-dependence of $q_{\rm d}$ via gas plus dust simulations, we now turn to the use of analytical calculations, which are presented in the next section.

   % -------------------
   \section{Semi-analytical formula for the pebble isolation mass}
   \label{semiformula}
   % -------------------
   In the previous section we showed that dust turbulent diffusion could significantly alter the PIM that is inferred from the necessary condition to form a radial pressure bump beyond the  orbit of the planet. As dust turbulent diffusion depends on the Stokes number (in addition to the gas turbulent viscosity), the PIM should explicitly depend on the Stokes number (in addition to the gas turbulent viscosity and aspect ratio). To examine how the PIM depends on the Stokes number, we turn to analytical and semi-analytical calculations. The idea is  that particles get trapped at the pressure bump outside th orbit of the planet if the radial drift velocity ($v_{\mathrm{drift}}$) of the particles inside of the pressure bump, which is positive, becomes larger than the root mean square turbulent velocity fluctuations ($v_{\mathrm{turb}}$). When such a condition is satisfied, the particles that are kicked into the gap by the turbulence drift back towards the pressure maximum. We use Eq.~(37) of \cite{2007Icar..192..588Y} to calculatie the radial profile of $v_{\mathrm{turb}}$, and their Eqs.~(57) and (59) for that of $v_{\mathrm{drift}}$. These equations are written as
   \begin{eqnarray}
     v_{\mathrm{turb}} &=& \sqrt{\frac{1+4\mathit{St}^2}{(1+\mathit{St}^2)^2} \alpha} h v_{\rm K},\label{v_turb}\\
   v_{\mathrm{drift}} &=& \frac{\mathit{St}}{1+\mathit{St}^2} \frac{\partial \log P}{\partial \log r} h^2 v_{\rm K} \label{v_drift}.  
   \end{eqnarray}

  Equating these two velocities gives
	\begin{equation}
		\frac{\partial \log P}{\partial \log r}\bigg|_{v_{\mathrm{turb}} = v_{\mathrm{drift}}} = \frac{\sqrt{\alpha}}{h} \sqrt{\frac{1}{\mathit{St}^2}+4}.
		\label{dlnpdlnrbalance}
	\end{equation}	   
   This is the minimum (positive) pressure gradient required in the gap to keep the particles with Stokes number $\mathit{St}$ from diffusing inwards through the gap. Eq.~(\ref{dlnpdlnrbalance}) shows that for $\mathit{St} \gg 1$, $\partial \log P/ \partial \log r$ tends to $2 \sqrt{\alpha}/h$, independently of $\mathit{St}$, while it tends to $\sqrt{\alpha}/(\mathit{St}\ h)$ for $\mathit{St} \ll 1$. We thus expect the PIM to change behaviour when varying $\mathit{St}$ around $\mathit{St} = 1$, as shown below.
   
   To find the planet mass that produces the pressure gradient satisfying Eq.~(\ref{dlnpdlnrbalance}) in the gap, we 
   need to know how the planet modifies the gas pressure profile, or actually the gas density profile, in a locally isothermal disk 
   model (where the planet does not alter the gas temperature). We use the density profile given by Eqs.~(12)-(16) of 
   \cite{2015ApJ...807L..11D} which is, overall, in fairly good agreement with the results of disk-planet simulations for partial 
   gap-opening planets; we return to this point later in this section. This density profile can be written 
   as $\Sigma(r)=\Sigma_{0}(r)\,C_{\Sigma}(r)$, 
   where $\Sigma_0$ denotes the initial (unperturbed) density profile, and $C_{\Sigma}$ is a function of 
   $q$, $h$, $\alpha,$ and  $r$ through an integral function that depends on $s = - \partial \log \Sigma_{0}/\partial \log r$ 
   and $f = \partial \log h/\partial \log r$. Assuming a steady-state viscous disk model, where the mass accretion rate 
   $\dot{M}=3\pi \nu \Sigma_0$ is uniform, we have $s=1/2 + 2f$ (since $\alpha$ is taken uniform), which implies that 
  \begin{equation}
     \frac{\partial \log P}{\partial \log r} = -\frac{3}{2} + \frac{\partial \log C_{\Sigma}}{\partial \log r}.
     \label{modifieddpdr2}
   \end{equation} 
   Anticipating that $C_{\Sigma}(r)$ and the PIM have a weak dependence on the disk's flaring index $f$, as is shown below, we take $f=0$ and $s=1/2$, as in our hydrodynamical simulations. Because $C_{\Sigma}(r)$ does not have an easily tractable form, we use an iterative method to find the right planet-to-star mass ratio $q$ such that Eqs.~(\ref{dlnpdlnrbalance}) and~(\ref{modifieddpdr2}) are satisfied, that is such that
   \begin{equation}
   \frac{\partial \log C_{\Sigma}}{\partial \log r} = \frac{3}{2} +  \frac{\sqrt{\alpha}}{h} \sqrt{\frac{1}{\mathit{St}^2}+4}.
   \label{goal}
   \end{equation}   
   For each pair $\{h,\alpha\}$, we first find the planet-to-star mass ratio $q_{\rm g,A}$ such that a radial pressure maximum builds up beyond the orbit of the planet ($\partial\log C_{\Sigma} /\partial\log r \approx 3/2$ from Eq.~\ref{modifieddpdr2}). Then, for each Stokes number $\mathit{St}$, we increase the planet-to-star mass ratio from $q=q_{\rm g,A}$ until Eq.~(\ref{goal}) is satisfied in the gap. The planet-to-star mass ratio that we obtain, which we denote by $q_{\rm d,A}$, corresponds to the PIM-to-star mass ratio for the {\it effective} trapping of particles with Stokes number $\geq\mathit{St}$ beyond the planet gap. The subscript {\it A} in $q_{\rm g,A}$ and $q_{\rm d,A}$ indicates that these quantities are calculated using our semi-analytical method.
   
   % FFFFFFFFFFFF
   \begin{figure}%[!h]
   \centering
   \includegraphics[width=\hsize]{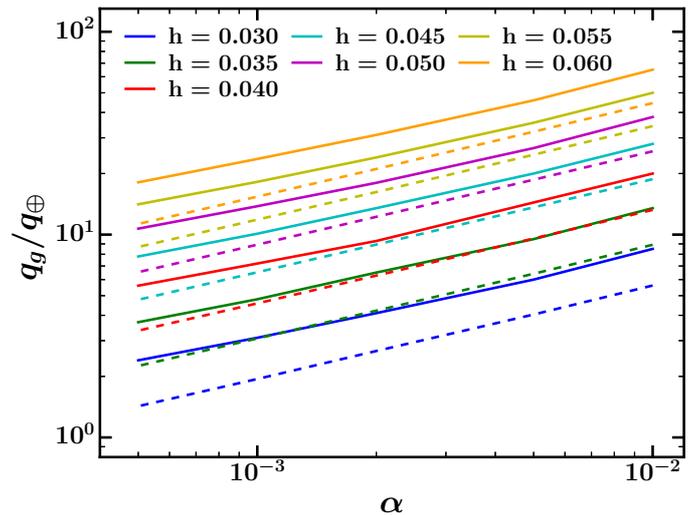}     
   \caption{Pebble isolation mass (normalized to the Earth mass) without effects of dust turbulent diffusion. A comparison between our results of gas-only hydrodynamical simulations (solid curves, Eq.~\ref{gasfitformula}) with the prediction of the analytical gap model of \cite{2015ApJ...807L..11D} (dashed curves, Eq.~\ref{gasfitformulaDuffell}).}
   \label{qgfomula}
   \end{figure}
 % FFFFFFFFFFFF
   
The calculation of $q_{\rm g,A}$ with the analytic density profile of \cite{2015ApJ...807L..11D} is well approximated by
 \begin{equation}
      \left(\frac{q_{\rm g,A}}{h^{3}}\right)^2 \approx 37.7\alpha+0.01.
   \label{gasfitformulaDuffell}
   \end{equation}
   By comparison with Eq.~(\ref{gasfitformula}), we see that for each pair $\{h,\alpha\}$, $q_{\rm g,A}$ is smaller than its value in our hydrodynamical simulations by a factor $\approx 1.5$. This comparison is illustrated in Fig.~\ref{qgfomula}.
   
   % FFFFFFFFFFFF
   \begin{figure}
   \centering
     \includegraphics[width=\hsize]{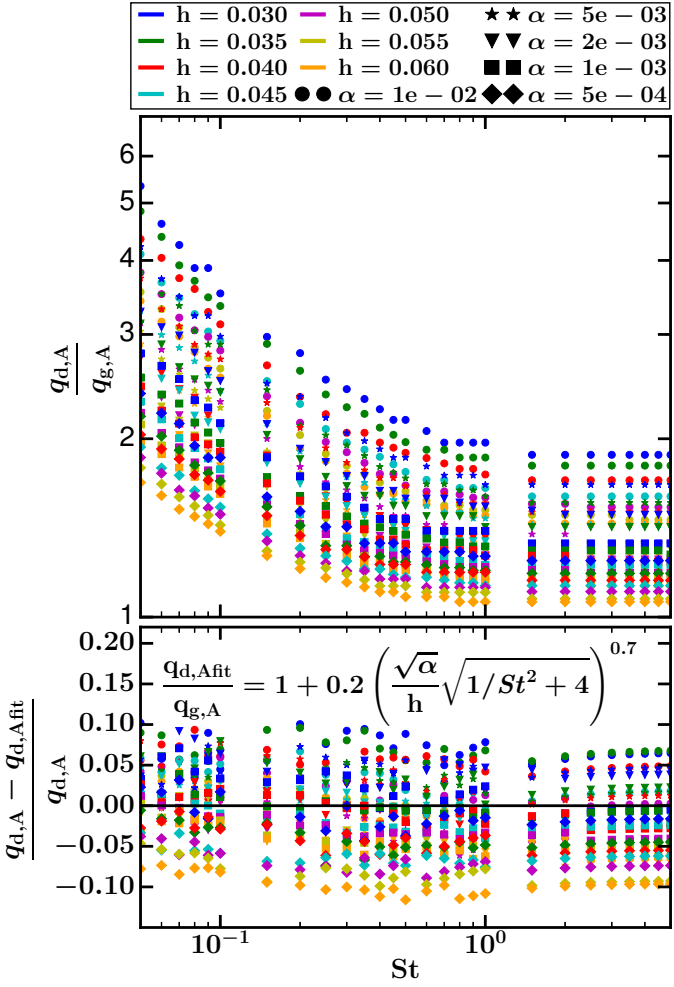}
     \caption{Semi-analytical calculation showing the dependence of the PIM on dust turbulent diffusion, using the method in Sect.~\ref{semiformula}. {\it Upper panel}: Ratio between the normalized PIM with dust turbulent diffusion ($q_{\rm d,A}$) and that without ($q_{\rm g,A}$), as a function of the particles Stokes number. Both $q_{\rm g,A}$ and $q_{\rm d,A}$ are calculated using the analytical gap density profile of \cite{2015ApJ...807L..11D}. {\it Lower  panel}: Relative difference between the normalized PIM obtained with our semi-analytical calculation ($q_{\rm d,a}$) and that obtained with our fitting formula ($q_{\rm d,A\,fit}$, Eq.~\ref{dustfitformula}).}
     \label{Stdipendence}
   \end{figure}   
   % FFFFFFFFFFFF
  
   Fig.~\ref{Stdipendence} shows the results of our semi-analytical calculation of the ratio $q_{\rm d,A}/q_{\rm g,A}$. For the range of Stokes numbers, aspect ratios, and alpha turbulent viscosities that we explored, we obtain good agreement with the following expression:
   \begin{equation} 
      \frac{q_{\rm d,A}}{q_{\rm g,A}} \approx 1+ 0.20 \left(\frac{\sqrt{\alpha}}{h}\sqrt{\frac{1}{ \mathit{St}^2}+4} \right)^{0.7},
      \label{dustfitformula}
   \end{equation}
   which, from Eq.~(\ref{dlnpdlnrbalance}), can be recast more simply as
    \begin{equation}
       \frac{q_{\rm d,A}}{q_{\rm g,A}} \approx 1+ 0.20 \left(\frac{\partial \log P}{\partial \log r} \right)^{0.7}_{v_{\mathrm{turb}} = v_{\mathrm{drift}}}.
       \label{dustfitformula1}
   \end{equation}
   As shown by the lower panel in Fig.~\ref{Stdipendence}, the agreement between Eq.~(\ref{dustfitformula}) and our results of semi-analytical calculations is within $\approx$10\%. In particular, we note that for a given pair $\{h,\alpha\}$, $q_{\rm d,A}/q_{\rm g,A}$ decreases with increasing $\mathit{St}$ for $\mathit{St} \ll 1$, and progressively becomes independent of $\mathit{St}$ for $\mathit{St} \gtrsim 1$, in agreement with the qualitative behaviour of $\partial\log P / \partial\log r$ given by Eq.~(\ref{dlnpdlnrbalance}) and discussed above. We finally stress that it is not entirely surprising that $q_{\rm d,A}/q_{\rm g,A}$ features the pressure gradient given by Eq.~(\ref{dlnpdlnrbalance}), since it is the parameter that sets the effective trapping of particles with a given Stokes number in the presence of dust turbulent diffusion.
      
   % FFFFFFFFFFFF
   \begin{figure}
   \centering
     \includegraphics[width=\hsize]{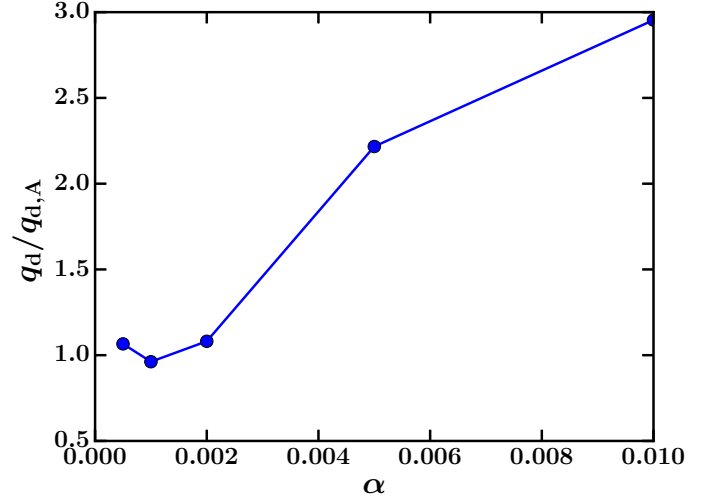}
     \caption{Ratio of the normalized PIM for $\mathit{St} \mathbf{\geq} 0.05$ particles obtained in our gas plus dust hydrodynamical simulations ($q_{\rm d}$) and by the semi-analytical calculation given by Eq.~(\ref{dustfitformula}) ($q_{\rm d,A}$). Results are for a disk aspect ratio $h=0.04$ and for all our $\alpha$ values.}
     \label{dustdist}
   \end{figure}   
  % FFFFFFFFFFFF
  Fig.~\ref{dustdist} compares the PIM inferred from the gas plus dust hydrodynamical simulations presented in Sect.~\ref{dustsimsresults} with the PIM obtained with the above semi-analytical calculation, and expressed via Eq.~(\ref{dustfitformula}); we denote by $q_{\rm
d}$ its value normalized to the mass of the central star. Results are for $h=0.04$ and $\mathit{St}  \mathbf{\geq} 0.05$ particles. We see that $q_{\rm d}$ and $q_{\rm d,A}$ are in good agreement overall, although $q_{\rm d}$ tends to be larger than $q_{\rm d,A}$ at large viscosities. For example, for $\alpha=10^{-2}$, $q_{\rm d}$ is about 2.9 times larger than $q_{\rm d,A}$. 
   
   % FFFFFFFFFFFF
   \begin{figure}%[!h]
   \centering
   \includegraphics[width=\hsize]{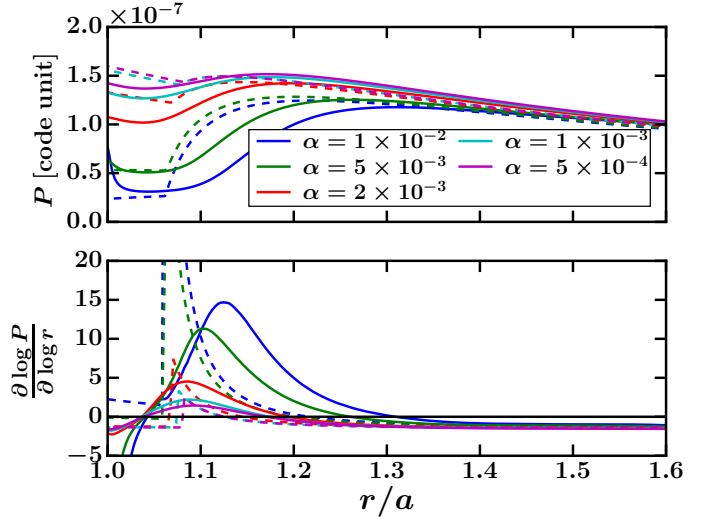}     
   \caption{Radial profiles of the gas pressure (upper panel) and of the logarithmic radial pressure gradient (lower panel) from our hydrodynamical simulations (solid curves) and from the analytic gap model of \citet{2015ApJ...807L..11D} (dashed curves).}
   \label{dpdrformula}
   \end{figure}
   % FFFFFFFFFFFF
   The difference between $q_{\rm d}$ and $q_{\rm d,A}$ can be due to the discrepancy between the analytical gap density profile of \citet{2015ApJ...807L..11D} and that in our hydrodynamical simulations. To investigate this, we show in Fig.~\ref{dpdrformula} the radial profiles of the gas pressure (upper panel) and the logarithmic radial pressure gradient (lower panel) obtained from our simulations (solid curves) and from \citet{2015ApJ...807L..11D} (dashed curves). Curves are shown for various alpha turbulent viscosities, but for the same planet mass and disk aspect ratio. For $\alpha \in [5\times10^{-4} - 2\times10^{-3}]$, there is very good agreement in both the maximum value of $\partial\log P / \partial\log r$ and its radial location, which explains why $q_{\rm d} / q_{\rm d,A} \approx 1$ for these values of $\alpha$. For $\alpha=5\times 10^{-3}$ and $10^{-2}$, however, the agreement is not so good, the maximum in $\partial\log P / \partial\log r$ being much larger and located closer to the planet in the analytical profile than in the simulations.  Consequently, a smaller planet mass is needed to satisfy the condition $v_{\rm drift} \simeq v_{\rm turb}$ than the PIMs obtained in the simulations, which explains why $q_{\rm d} / q_{\rm d,A} > 1$ for these values of $\alpha$. Based on this comparison, we thus suggest that Eq.~(\ref{dustfitformula}) should be considered as a lower limit for the PIM in high-viscosity disks, and a good estimate for the PIM in low-viscosity disks.
     
\begin{figure}%[!h]
   \centering
   \includegraphics[width=\hsize]{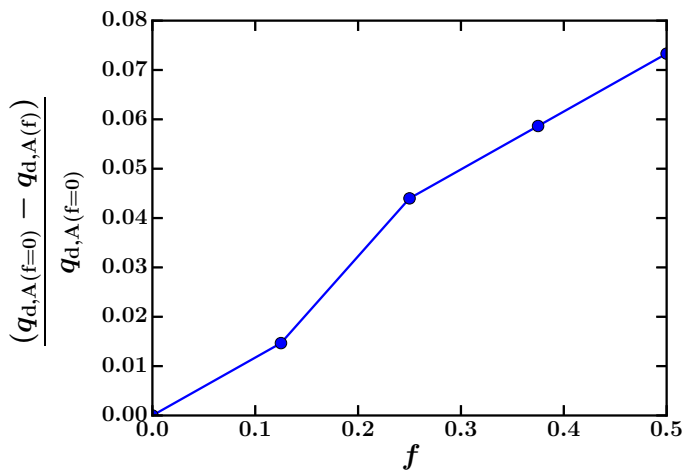}     
   \caption{Dependence of the PIM on the disk's flaring index $f$, as obtained from our semi-analytical calculation for a disk aspect ratio $h=0.04$, turbulent viscosity $\alpha = 10^{-3}$ and for particles Stokes number $\mathit{St} = 0.1$.}
   \label{qdflaring}
   \end{figure}

The expressions of $q_{\rm g,A}$ and $q_{\rm d,A}$ in Eqs.~(\ref{gasfitformulaDuffell}) and~(\ref{dustfitformula}) are for a steady-state viscous disk model with flaring index $f=0$. As stated above, the background pressure profile scales as $r^{-3/2}$ for a steady-state viscous disk model, whatever the specific choice of density and temperature profiles. However, the perturbed gap density profile ($C_{\Sigma}$ in Eq.~\ref{modifieddpdr2}) depends on the gradients of the background density and temperature profiles in a way that does not feature the background pressure gradient. The PIM should therefore also depend on the background temperature gradient (or, equivalently, on the surface density gradient, since both quantities are related in a steady-state viscous disk model). We checked with our semi-analytical calculation that this dependence remains weak. This is illustrated in Fig.~\ref{qdflaring}, which shows that $q_{\rm d,A}$ slightly decreases with increasing flaring index. This decrease is by $\approx$7\% for $f=0.5$, which corresponds to a uniform background temperature. We also verified this weak dependence by running gas plus dust simulations for a radiative disk model with background gas surface density in $r^{-15/14}$ and temperature in $r^{-3/7}$ \citep[e.g.][]{2014A&A...564A.135B}, for which we find basically the same PIM as with our fiducial disk profiles.

   % FFFFFFFFFFFF
   % FFFFFFFFFFFF
   
   % ===============
   \section{Summary}
   % ===============
   \label{summary}
   In this work we have examined how disk turbulence affects the PIM. Our findings can be summarized as follows:
   \begin{enumerate}
      \item First, we carried out gas hydrodynamical simulations of planet-disk interactions to find the minimum planet mass to build up a radial pressure bump outside the orbit of the planet, as a function of disk aspect ratio $h$ and $\alpha$-turbulent viscosity. The formation of such a pressure bump is a necessary but not sufficient condition for trapping pebbles beyond the orbit of the planet. By denoting $q_{\rm g}$ the ratio between the PIM and the mass of the central star, our results of simulations are best reproduced by the following expression: $(q_{\rm g} / {h^3} )^2 \approx 82.33 \alpha + 0.03$. Planets that reach the PIM deplete the surface density in the disk gas around their orbit by 10 to 20\%. Our semi-analytical calculation based on the gap density profile of \citet{2015ApJ...807L..11D} gives a PIM-to-star mass ratio $q_{\rm g,A}$ such that $(q_{\rm g,A} / {h^{3}} )^2 \approx 37.3 \alpha + 0.01$. The difference between $q_{\rm g}$ and $q_{\rm g.A}$ resides in a slightly steeper gap pressure profile in \citet{2015ApJ...807L..11D} than in our hydrodynamical simulations.
      \medskip     
      \item In addition to the presence of a radial pressure maximum, particle trapping requires that the drift velocity remains larger than the root mean square turbulent velocity fluctuations in the gap. This sets a minimum pressure gradient in the gap, which is given by Eq.~(\ref{dlnpdlnrbalance}). Our semi-analytical calculation shows that with dust turbulent diffusion, the effective trapping of particles with Stokes number $\geq \mathit{St}$ gives rise to a correction factor by which $q_{\rm g,A}$ should be multiplied to obtain the actual PIM-to-star mass ratio ($q_{\rm d,A}$). This correction factor is given by Eq.~(\ref{dustfitformula}). Our final expression for the PIM-to-star mass ratio is written      
      \begin{equation}
         q_{\rm d,A} \approx \underbrace{h^{3} \sqrt{37.3 \alpha + 0.01}}_{q_{\rm g,A}} \times 
         \left[
         1+ 0.2 \left(\frac{\sqrt{\alpha}}{h}\sqrt{\frac{1}{ \mathit{St}^2}+4} \right)^{0.7}
         \right].
         \label{eq:finalPIM}
      \end{equation}
      \smallskip
      \item We also carried out gas plus dust hydrodynamical simulations with a range of particle sizes and we find that the PIM given by Eq.~(\ref{eq:finalPIM}) is in good agreement with our results of simulations for low-viscosity disks. This PIM\ takes, however, smaller values than in the simulations for high-viscosity disks (by up to a factor $\sim$3 for $\alpha = 10^{-2}$).
   \end{enumerate}
   
On the simulation side, this work is based on 2D hydrodynamical simulations of disk-planet interactions for a non-migrating planet. Such simulations require the use of a softening length in the planet's gravitational potential, which can be seen as a crude way to mimic the effects of a finite vertical thickness. \cite{2012A&A...546A..18M}, who performed 2D simulations with a softening length slightly larger than ours ($\epsilon=0.6H$ instead of $0.4H$), found that for $\alpha=6 \times 10^{-3}$ and $h=0.056$ the PIM is about 50~Earth masses, while our Eq.~(\ref{gasfitformula}) gives $\sim$ 42~Earth masses. Another comparison can be made with \cite{2014A&A...572A..35L}, whose 3D simulations resulted in a PIM of $\sim$20~Earth masses for $\alpha=6 \times 10^{-3}$ and $h=0.05$, while we find $\sim$30~Earth masses. This comparison indicates that our choice of softening length gives overall good agreement with 3D simulations. Furthermore, if the migration of the planet is considered, a possible flux of pebbles coming from the disk inside the orbit of the planet may increase the PIM, depending on how the migration timescale and the pebbles radial drift timescale compare. This particular point requires a dedicated study.
   
     A study of the PIM based on 3D gas-only hydrodynamical simulations along with 2D integrations of particle trajectories in the disk midplane has been recently undergone by \cite{2018arXiv180102341B}. A detailed comparison of our results with those of \cite{2018arXiv180102341B} is found in Appendix~\ref{App-bitsch} . 
   
    On the analytical side, our semi-analytical calculation of the PIM discards dust-gas interactions in the spirals and the direct gravitational interaction of the planet on the dust particles. Both of these effects could alter the predicted PIMs.

   The stability of the dust ring that forms by piling up and growth of the particles should also be investigated in future works. If the lifetime of the ring is short compared to the age of the disk, the observed ring in millimetre observations would need more massive planets to sustain a stronger pressure bump. On the other hand, the streaming instability might lead to a periodic growth of the planet by frequent destruction of the outer pressure bump. Therefore, studying the long-term evolution of dust trapping at the pressure maximum of planets with masses about the PIM and in low-viscosity disks (such that the low diffusion allows dust-to-gas ratios to reach unity) might probably need to take into account effects of dust drag onto the gas. Dust fragmentation and growth, which also depend on the disk turbulent diffusion, could also alter the PIM. These effects should be examined in future studies.
   
   % AAAAAAAAAAAAA
   \begin{acknowledgements}
    This work has been carried out within the frame of the National Centre
for Competence in Research PlanetS supported by the Swiss National Science Foundation. We are grateful to the referee for her/his constructive report that helped clarify this paper. We would also like to thank Paola Pinilla for her useful comments.
   \end{acknowledgements}

\appendix
% ===============
\section{Comparison with \cite{2018arXiv180102341B}}
% ===============
\label{App-bitsch}
  Very recently, \cite{2018arXiv180102341B} have studied the PIM dependency on the disk aspect ratio $h$, the $\alpha$ turbulent viscosity, and the initial pressure gradient $\partial \log P_0/\partial \log r$ using gas-only 3D hydrodynamical simulations. Upon integrating the trajectory of particles with different Stokes numbers in the midplane of the disk, these authors have shown that their PIM could indeed trap particles with $\mathit{St} \geq 0.01$ at the outer edge of the gap of the planet in the absence of dust turbulent diffusion. Using analytical estimations, they have obtained a correction factor that should be applied to their PIM expression to account for the effects of dust turbulent diffusion. Because the PIM expression obtained in \cite{2018arXiv180102341B} can be seen as the 3D counterpart to that of the present study, we compare here both PIM formulae.
  
  To ease the comparison, we express the formula of \cite{2018arXiv180102341B} with our notations. We denote  by $q_{\rm g3d}$ their PIM-to-star mass ratio without dust turbulent diffusion, and by $q_{\rm d3d}$ that obtained when accounting for dust turbulent diffusion. Eqs.~(10) and (11) of \cite{2018arXiv180102341B} can be recast as
  \begin{equation}
  	q_{\rm g3d} = 25 q_{\oplus} f_{\rm fit},
  	\label{bitschqg}
  \end{equation}
  where
  \begin{equation}
  	f_{\rm fit} = \left(\frac{h}{0.05}\right)^3 \left[ 0.34 \left(\frac{\log_{10} \alpha_{3}}{\log_{10} \alpha}\right)^4 + 0.66 \right] \left[\frac{5.5+s-f}{6}\right],
  	\label{ffit}
  \end{equation}
  with $\alpha_{3}=10^{-3}$.
        In calculating $\partial \log P/\partial \log r$ in Eq.~(11) of \cite{2018arXiv180102341B}, we use $P=\rho c_{s}^2$ with $\rho=\Sigma_0/(\sqrt{2\pi}H)$.
   
   % FFFFFFFFFFFF
   \begin{figure}
   \centering
     \includegraphics[width=\hsize]{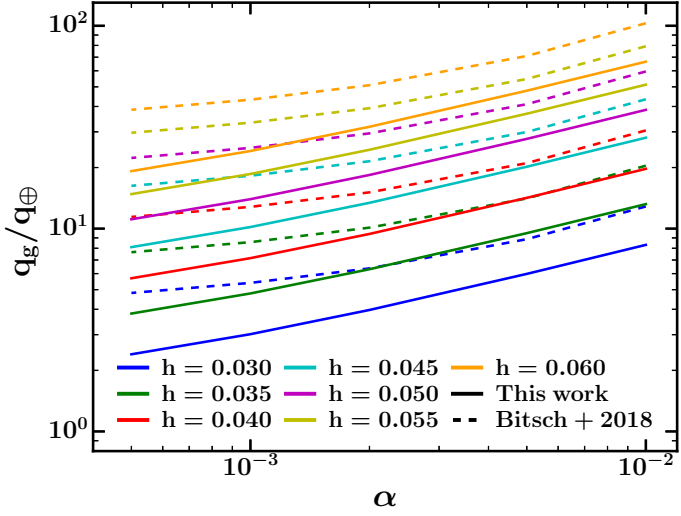}
     \caption{PIM (in Earth masses) without effects of dust turbulent diffusion. A comparison between the results of our 2D hydrodynamical simulations (solid curves) and the 3D simulations of \cite{2018arXiv180102341B} (dashed curves) for $\Sigma \propto r^{-0.5}$ and uniform aspect ratios.}
     \label{qgSB}
   \end{figure}   
  % FFFFFFFFFFFF

  In Fig.~\ref{qgSB}, we compare $q_{\rm g}$ as obtained in our 2D simulations with $q_{\rm g3d}$ from the 3D simulations of \cite{2018arXiv180102341B}. Both of these show overall good agreement and similar behaviours when varying $h$ and $\alpha$. Our PIM values are a factor 1.5 to 2 times smaller than in the 3D simulations of \cite{2018arXiv180102341B}. As shown in \cite{2018arXiv180102341B}, this can be understood by a more difficult gap formation in a 3D disk model than in 2D. \cite{2018arXiv180102341B} have also carried out a few 2D simulations, which are directly comparable to ours. For a background surface density profile in $ r^{-1/2}$, $h=0.05$, and $\alpha=10^{-3}$, their Fig.~B1 shows a PIM of about $17 M_{\oplus}$, while we obtain $14 M_{\oplus}$ in our 2D simulations. While this is an overall good agreement, the $20\%$ relative difference can be due to different grid resolutions (we use about twice as many grid cells in the radial direction), or perhaps to different simulation times (which are not specified in \citealp{2018arXiv180102341B}). As we have experienced during preliminary tests, the gap profile can evolve on long timescales (e.g. a few thousand planet orbits for $\alpha = 10^{-3}$), and running times that are too short lead to overestimating the PIM.
  
  To account for the effects of dust turbulent diffusion on the PIM, \cite{2018arXiv180102341B} have used a similar strategy as ours, in the sense that they also look for the planet mass in their simulations from which the root mean square turbulent velocity fluctuations of the particles ($v_{\rm turb}$) become equal to their radial drift velocity ($v_{\rm drift}$) within the gap. For the radial drift velocity, they have used
   \begin{equation}
  v_{\rm drift} = \mathit{St} \frac{\partial \log P}{\partial \log r} h^2 v_{\rm K},
   \label{v_driftB}
   \end{equation}
    which coincides with Eq.~(\ref{v_drift}) in the limit $\mathit{St} \ll 1$. For the root mean square turbulent velocity fluctuations, instead of Eq.~(\ref{v_turb}), \cite{2018arXiv180102341B} have used \begin{equation}
 v_{\rm turb} \sim D_{\rm d} / H_{\rm d}, 
 \label{v_turbB}
 \end{equation}
 with $H_{\rm d}$ a radial length scale for particle trapping in the pressure maximum, which is taken equal to the gas pressure scale height $H$, and with the dust's turbulent diffusion $D_{\rm d}$ taken equal to the gas kinematic viscosity $\nu$. We note that our expression for $D_{\rm d}$ tends to $\nu$ only in the limit $\mathit{St} \ll 1$. Also, we stress that in the same limit $\mathit{St} \ll 1$, Eq.~(\ref{v_turbB}) gives $v_{\rm turb} \approx \alpha c_{\rm s}$, while our Eq.~(\ref{v_turb}) gives $v_{\rm turb} \approx \sqrt{\alpha} c_{\rm s}$. Eq.~(\ref{v_turb}) thus leads to (much) larger turbulent velocity fluctuations than Eq.~(\ref{v_turbB}). Using Eqs.~(\ref{v_driftB}) and~(\ref{v_turbB}), the pressure gradient in the gap such that $v_{\rm turb} = v_{\rm drift}$ is
 \begin{equation}
 \frac{\partial \log P}{\partial\log r}\bigg|_{v_{\mathrm{turb}} = v_{\mathrm{drift}}} = \frac{\alpha} {h} \frac{1} {\mathit{St}},
 \label{criterionB}
 \end{equation}
 which is a factor $\alpha^{-1/2}$ smaller than our expression given by Eq.~(\ref{dlnpdlnrbalance}) in the limit $\mathit{St} \ll 1$, due to the different expressions for $v_{\rm turb}$ in that limit.
 
    Using Eq.~(\ref{criterionB}), \cite{2018arXiv180102341B} have found that the PIM with dust turbulent diffusion is written 
    \begin{equation}
        \frac{q_{\rm d3d}}{q_{\rm g3d}} \approx 1+4.2\frac{\alpha}{\mathit{St}},
        \label{qdBitsch}
    \end{equation}
    which, using Eq.~\ref{criterionB}, can be recast as
    \begin{equation}
        \frac{q_{\rm d3d}}{q_{\rm g3d}}  \approx 1 + 0.21 \left(\frac{h}{0.05} \right) \frac{\partial \log P}{\partial \log r}\bigg|_{v_{\mathrm{turb}} = v_{\mathrm{drift}}}.    
        \label{qdBitsch1}
    \end{equation}
    Interestingly, Eq.~(\ref{qdBitsch1}) formally resembles our expression given by Eq.~(\ref{dustfitformula1}), except mainly for the scaling with the pressure gradient (which in our case comes to the 0.7 power). However, because we have different expressions for the pressure gradient $\partial\log P/\partial\log r$, our PIM values should differ {\it a priori}.
    
\begin{figure}
   \centering
     \includegraphics[width=\hsize]{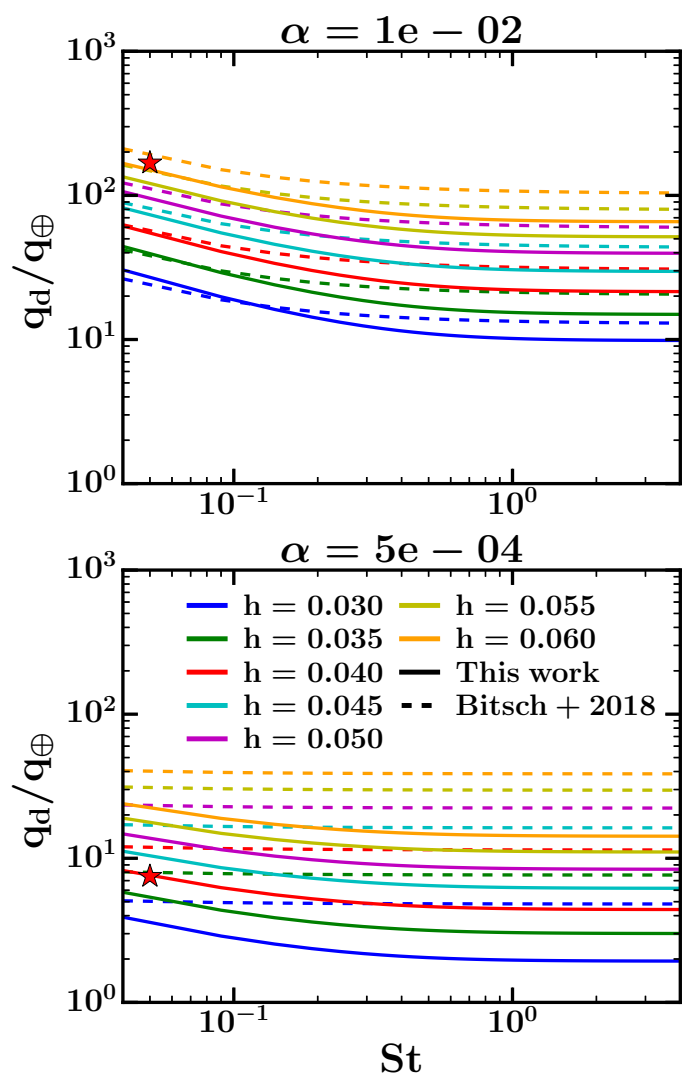}
     \caption{PIM (in Earth masses) with effects of dust turbulent diffusion. A comparison between the results of our semi-analytical calculation (solid curves) and those of \cite{2018arXiv180102341B} (dashed curves) for $\Sigma \propto r^{-0.5}$, uniform aspect ratios and two values of the turbulent viscosity of the disk: $\alpha = 10^{-2}$ (upper panel) and $\alpha = 5\times 10^{-4}$ (lower panel). The red stars indicate the PIM obtained from our 2D gas plus dust simulations (for $h=0.04$).}
     \label{qdSB}
   \end{figure}

     Fig.~\ref{qdSB} compares $q_{\rm d3d}$ (see Eqs.~\ref{qdBitsch}, ~\ref{bitschqg}, and~\ref{ffit}) with our expression for $q_{\rm d}$ (see Eq.~\ref{eq:finalPIM}) for $\alpha = 10^{-2}$ (upper panel) and $\alpha = 5\times 10^{-4}$ (lower panel). For the regime of Stokes numbers that we are interested in, our PIM with dust turbulent diffusion is overall in good agreement with \cite{2018arXiv180102341B} for $\alpha = 10^{-2}$. However, for $\alpha=5 \times 10^{-4}$, we predicted smaller PIMs than \cite{2018arXiv180102341B}, for which the PIM displays no dependency on Stokes number for that viscosity. The differences are due to the combined effects of (i) different prescriptions for calculating the particle velocities (particularly the turbulent velocity fluctuations, as shown above), (ii) different dependencies of the PIM with the pressure gradient (compare Eqs.~\ref{qdBitsch1} and \ref{dustfitformula1}), and also (iii) a larger $q_{g}$ in \cite{2018arXiv180102341B}.
    
  A last point of comparison with \cite{2018arXiv180102341B} is the effect of varying the initial density and temperature gradients on the PIM. While we have specialized to steady-state viscous disk models where both gradients are related, \cite{2018arXiv180102341B} have varied both gradients independently. In Fig.~\ref{qdflaring}, we show with our semi-analytical method that the PIM slightly decreases with increasing the background temperature gradient $\partial\log T / \partial\log r$. \cite{2018arXiv180102341B} have found with their simulations that the PIM slightly decreases with increasing the initial pressure gradient, which looks consistent with our findings, except that this dependence is found with increasing the initial density gradient for a fixed temperature profile ($f=0$, see their Fig. 3), whereas no dependence of the PIM is found when varying the initial temperature gradient for a fixed surface density profile ($s=1/2$, see their Sect. 2.5). The reason for this is not totally clear, and may have to do with the steady-state assumption in our disk models. As stated before, starting with a non-steady-state disk model implies larger simulation times for a proper estimate of the PIM.
  
   % FFFFFFFFFFFF
  % FFFFFFFFFFFF  
   
   \bibliographystyle{aa}

\end{document}